# Optimal Design of Waveform Digitisers for Both Energy Resolution and Pulse Shape Discrimination

Jirong Cang [a,b], Tao Xue [a,b], Ming Zeng [a,b*], Zhi Zeng [a,b], Hao Ma[a,b], Jianping Cheng [a,b] and Yinong Liu[a,b]

[a] *Key Laboratory of Particle & Radiation Imaging(Tsinghua University), Ministry of Education, China*

[b] *Department of Engineering Physics, Tsinghua University, Beijing 100084, China*

## Abstract

Fast digitisers and digital pulse processing have been widely used for spectral application and pulse shape discrimination (PSD) owing to their advantages in terms of compactness, higher trigger rates, offline analysis, etc. Meanwhile, the noise of readout electronics is usually trivial for organic, plastic, or liquid scintillator with PSD ability because of their poor intrinsic energy resolution. However, $LaBr_3$(Ce) has been widely used for its excellent energy resolution and has been proven to have PSD ability for alpha/gamma particles. Therefore, designing a digital acquisition system for such scintillators as $LaBr_3$(Ce) with both optimal energy resolution and promising PSD ability is worthwhile. Several experimental research studies about the choice of digitiser properties for liquid scintillators have already been conducted in terms of the sampling rate and vertical resolution. Quantitative analysis on the influence of waveform digitisers, that is, fast amplifier (optional), sampling rates, and vertical resolution, on both applications is still lacking. The present paper provides quantitative analysis of these factors and, hence, general rules about the optimal design of digitisers for both energy resolution and PSD application according to the noise analysis of time-variant gated charge integration.

## 1. Introduction

The $LaBr_3$(Ce) scintillator has been widely studied for gamma-ray spectroscopy owing to its excellent energy resolution (<3% at 662 keV), detection efficiency, and time resolution. In addition, $LaBr_3$(Ce) has been proven to have the ability for pulse shape discrimination (PSD) between alpha and gamma events [1, 2]. The PSD ability extends the application of $LaBr_3$(Ce) for low-activity measurement by distinguishing the alpha contamination from $^{227}$Ac (energy > 1.6 MeV). Consequently, the design of an acquisition system for $LaBr_3$(Ce) with simultaneous optimal energy resolution and promising PSD ability is worthwhile.

With regards to the PSD, it has been significantly improved owing to the development of fast digitisers during the past years. Several research studies on the influence of digitisers on the PSD performance in organic [3-5], plastic [6], or liquid [7, 8] scintillators have been conducted. Recently, McFee et al. [9] studied the PSD performance and energy spectrum application using digitised lanthanum halide scintillator pulses. However, optimal results were not achieved, which was assumed to be due to the degradation of the digitiser. Therefore, research works through quantitative analysis of the influence of digitisers on the energy resolution and PSD remain limited, which should be necessary to provide a general rule about the optimal design of waveform digitisers for specific applications. The present study quantitatively analysed the influence of digitiser properties in terms of fast amplifiers, sampling rate ($F_S$), and vertical resolution on the spectral resolution and PSD performance. According to the quantitative analysis, sufficiently good energy resolution and PSD performance can be achieved using a moderate digitiser for $LaBr_3(Ce)$. Furthermore, this analysis also provides a general rule about the optimal design of waveform digitisers for similar applications based on digital waveform processing.

* Corresponding author, *E-mail*: zengming@tsinghua.edu.cn

## 2. System Model and Noise Analysis of Gated Integration

A general digital waveform sampling system for scintillators is shown in Fig. 1, where the scintillator is usually coupled to a photomultiplier tube (PMT). The digitiser consists of a fast amplifier, an analogue-to-digital converter (ADC), and other modules responsible for data transfer or storage. The fast amplifier functions as a signal-amplification and/or anti-aliasing filter, which tunes the input signal to obtain a better signal-to-noise ratio (SNR), whereas the ADC is responsible for the digitisation.

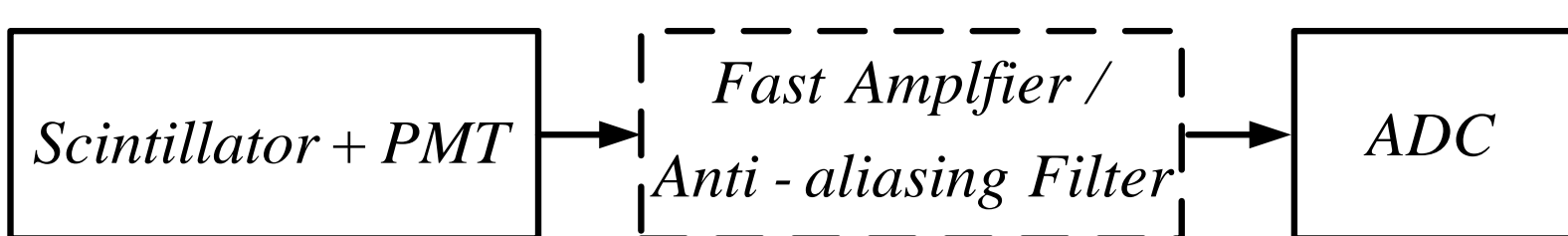

Fig. 1 Digital waveform sampling system for scintillators

In the following discussion, we will focus on the noise analysis of how the fast amplifier and ADC influence the energy resolution and PSD ability (the data transfer and storage are considered negligible).

Fig. 2 shows that the total integration (long gate) of the current waveform corresponds to the total energy deposited in the scintillator. The charge comparison method (CCM, also called gated integration) is most widely used in PSD and has been proven to be a good method for the PSD in a $LaBr_3(Ce)$ detector [1, 2]. The PSD feature can be expressed as

$$CCM = \frac{Q_p}{Q_t} = \frac{Short\ Gate\ Integration}{Long\ Gate\ Integration} \tag{1}$$

In this study, methods based on gated charge integration are quantitatively analysed, and hence, the influences of digitisers on both energy resolution and PSD application are classified.

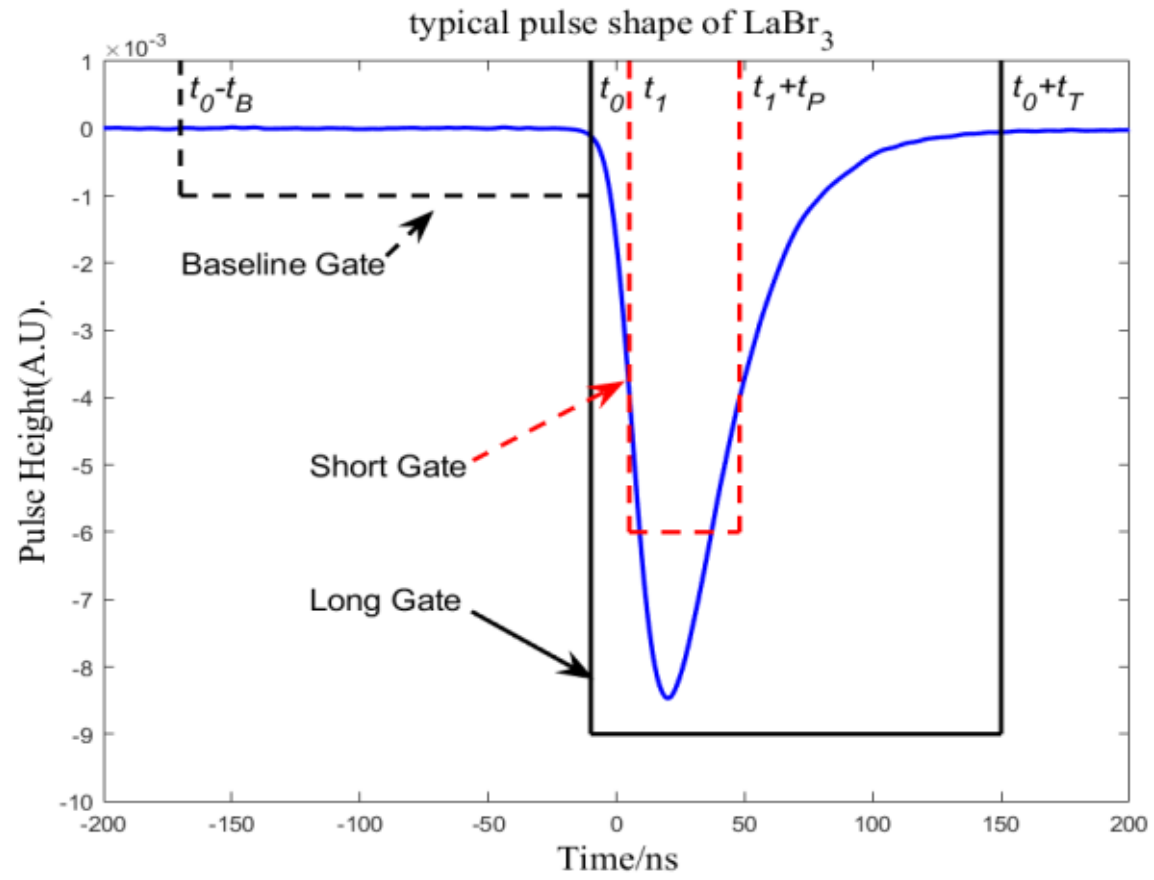

Fig. 2 Gated charge integration

*2.1 Noise Analysis of Gated Integration in a Simplified Analogue Domain*

Considering a traditional analogue integration, a simplified system response diagram that uses a one-order low-pass filter as a representation of a fast amplifier is shown in Fig. 3

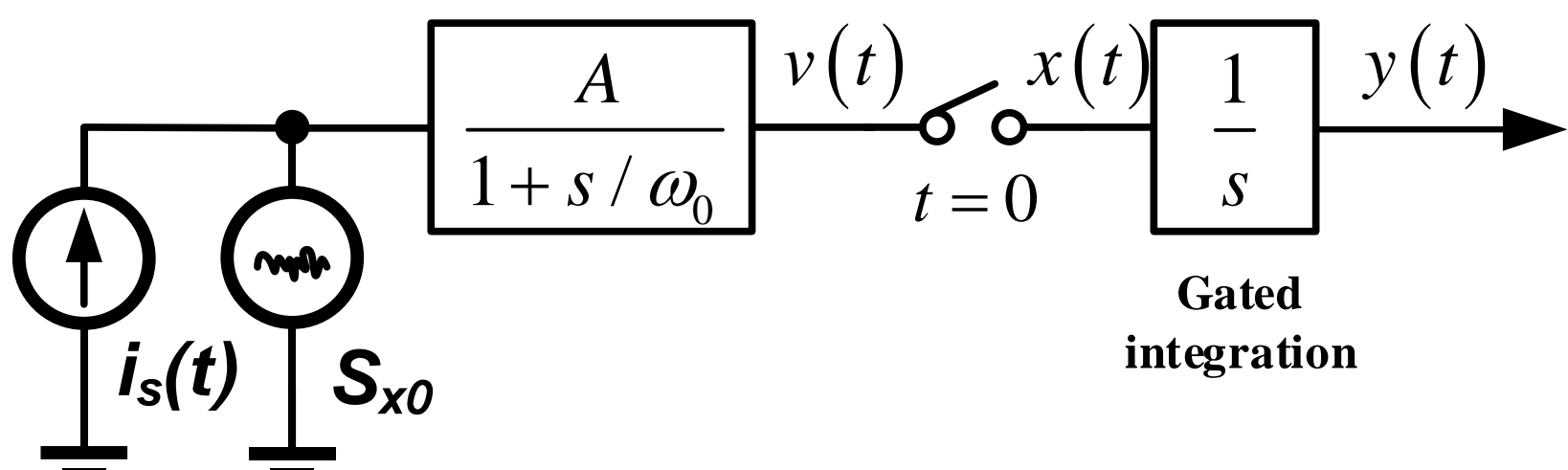

Fig. 3 Analogue diagram of gated charge integration

The measured gated charge integration is composed of the signal and noise of the digitisers, that is,

$$\begin{aligned} Q_m &= \int_0^t \left( x_s(\tau) + x_n(\tau) \right) d\tau \\ &= Q_s + Q_n \end{aligned} . \tag{2}$$

The detailed inference can be referred in Ref. [10]. As a simplified inference, the autocorrelation of the input noise is expressed as

$$R_{xx}(t_1, t_2) = A^2 \cdot S_{x0} \frac{\omega_0}{4} e^{-\omega_0 |t_1 - t_2|} . \tag{3}$$

The autocorrelation of the output noise of the gated integration can be expressed as

$$R_{yy}(t_1, t_2) = h(t_1) * R_{xx}(t_1, t_2) * h(t_2) , \tag{4}$$

where the impulse response of the gated integrator is $h(t) = 1$. Therefore, the uncertainty of the gated integration caused by the noise can be calculated as

$$
\begin{aligned}
\sigma_y^2(t) &= E\left[y(t)\,y(t)\right] = R_{yy}(t_1,t_2)\,|_{t_1=t_2=t} \\
&= A^2 \cdot \frac{S_{x0}}{2}\left(t - \frac{1}{\omega_0}\cdot\left(1-e^{-\omega_0 t}\right)\right)
\end{aligned}
\quad . \tag{5}
$$

If gated integration time is much bigger than the time constant of fast amplifier, that is $t \gg 1/\omega_0$, which is normally satisfied, the uncertainty of the integrated charge caused by the noise can be simplified as

$$
\sigma_y^2(t) \approx A^2 \cdot \frac{S_{x0}}{2} \cdot t \, . \tag{6}
$$

By normalising with current-to-voltage (*I–V*) gain *A*, the input-referred uncertainty of the gated charge caused by the noise is expressed as

$$
\sigma_{Q_n}^2(t) \approx \frac{1}{2} S_{x0} \cdot t \, . \tag{7}
$$

*2.2 Noise Analysis of the Gated Integration in a Digital Domain*

In the discrete digital domain, the analogue integration will be replaced by the sum of the digital data, and the system diagram is shown in Fig. 4.

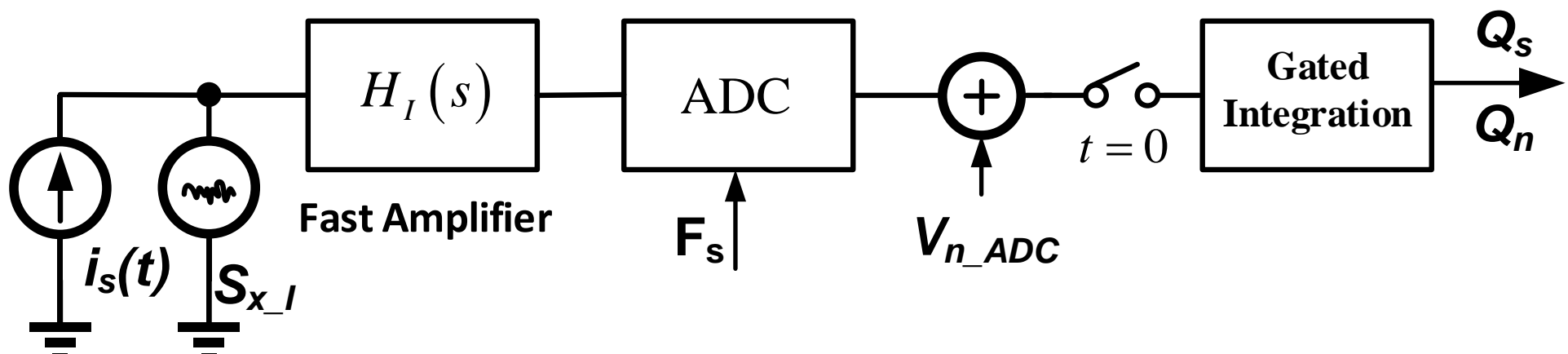

Fig. 4 Digital diagram of the gated charge integration

The measured gated charge integration can be expressed as

$$
\begin{aligned}
Q &= \sum\nolimits_k \left(i_s(k) + i_n(k)\right)\cdot T_s \\
&= Q_s + Q_n
\end{aligned}
\quad . \tag{8}
$$

The noise is usually considered as independent of the signals. Thus, the uncertainty of $Q_n$ can be expressed as

$$
\sigma_{Q_n}^2 = T_s^2 \cdot \sum\nolimits_j \sum\nolimits_k \operatorname{cov}\left[i_n(j), i_n(k)\right]. \tag{9}
$$

Fig. 4 shows that the noise mainly consists of two parts: fast amplifier and ADC noise. We need to mention that the noise from ADC can be normally considered uncorrelated in most situations, whereas the noise from the fast amplifier is usually band-limited, which is correlated. In the next two subsections, we will separately present their analysis, although we will end up with the same formula, that is, Eq. (7).

*Analysis of Uncorrelated Noise from ADC*

All ADC internal circuits produce a certain amount of broadband, that is, an uncorrelated noise due to the resistor and 'kT/C' noises, which is normally called *input-referred noise ($V_{ir}$)*. Meanwhile, some correlations exist between the quantisation error ($V_q$) and the input signal, especially if the input signal is an exact sub-multiple of the sampling frequency. In other words, the total noise of the ADC can be expressed as follows:

$$V_{n_ADC} = \sqrt{V_{ir}^2 + V_q^2} \,. \tag{10}$$

The quantisation error of an $N$-bit ADC can be expressed as follows, where $FUS$ means the full scale of the ADC.

$$V_q = \frac{FUS}{2^N} / \sqrt{12} \tag{11}$$

Moreover, the total noise ($V_{n_ADC}$) can be represented by another term called effective number of bits (ENOB), which is usually an important character of an ADC, that is,

$$V_{n_ADC} \approx \frac{FUS}{2^{ENOB}} / \sqrt{12} \,. \tag{12}$$

In most situations, not only the *input-referred noise* but also the quantisation error can be considered uncorrelated if the input-referred noise is larger than one-half of the least significant bit (LSB) [11] or an ac signal that spans more than a few LSBs [12]. Then, the covariance of the total noise from the ADC can be simplified as:

$$\operatorname{cov}\left[i_n(j), i_n(k)\right] = \begin{cases} V_{n_ADC}^2 / A^2, & j = k \\ 0 & , j \neq k \end{cases}, \tag{13}$$

where $A$ is the $I$–$V$ gain of the fast amplifier and ADC. Eq. (9) can be calculated as

$$\sigma_{Q_{n_ADC}}^2 = T_s^2 \cdot N \cdot \frac{V_{n_ADC}^2}{A^2} = \frac{V_{n_ADC}^2}{A^2 \cdot F_s} \cdot t \,. \tag{14}$$

Actually, the noise power spectral density of the ADC can usually be considered as a constant from DC to one-half of the sampling rate (Fs/2), which is

$$S_{x_ADC} = \frac{V_{n_ADC}^2}{A^2 \cdot F_s / 2} \,. \tag{15}$$

Then, Eq. (14) can be rewritten similar to Eq. (7) as

$$\sigma_{Q_{n_ADC}}^2 = \frac{1}{2} S_{x0_ADC} \cdot t \,. \tag{16}$$

*Analysis of Correlated Noise from the Fast Amplifier*

The noise from the fast amplifier is usually band-limited, which can be designed as the highest frequency of the input signal ($f_0$), that is, $\omega_0 = 2\pi f_0$. Here, we first use a one-order low-pass filter as the representation of the fast amplifier. By normalising with $I$–$V$ gain $A$, the system response of the fast amplifier can be simplified as

$$H_I(s)=\frac{1}{1+s/\omega_0}. \tag{17}$$

According to Eq. (3), the covariance of the noise at the $j$th and $k$th sampling times can be expressed as

$$\mathrm{cov}\left[i_n(j),i_n(k)\right]=S_{x_I}\cdot\frac{\omega_0}{4}\cdot e^{-\omega_0 T_s\cdot|j-k|}. \tag{18}$$

By substituting Eq. (18) to Eq. (9), the uncertainty of the gated integration of the noise from the fast amplifier is obtained as

$$\sigma^2_{Q_{n_I}}=S_{x_I}\cdot\frac{\omega_0}{4}\cdot T_s^2\cdot\left[\frac{t}{T_s}\cdot\frac{1+q}{1-q}-2\left(1-e^{-\omega_0 t}\right)\cdot\frac{q}{\left(1-q\right)^2}\right], \tag{19}$$

where $q=e^{-\omega_0 T_s}$ and $t$ is the time window of the gated integration. As an example, if $F_s\rightarrow\infty$, the limit of Eq. (19) will end up to be the same as that of Eq. (5) in the analogue domain.

$$\sigma^2_{Q_{n_I}}\left(F_s\rightarrow\infty\right)=\frac{S_{x_I}}{2}\cdot\left(t-\frac{1}{\omega_0}\cdot\left(1-e^{-\omega_0 t}\right)\right) \tag{20}$$

However, we need to understand in detail how the sampling rate influences the gated integration of the noise. According to Eq. (19), we can define a noise factor to demonstrate the influence of the sampling rate ($F_s$), that is,

$$NF\left(F_s\right)=\frac{\sigma_{Q_{n_I}}\left(F_s\right)}{\sigma_{Q_{n_I}}\left(F_s\rightarrow\infty\right)}. \tag{21}$$

As an illustration, two different filters (a first-order low-pass filter and a second-order Butterworth filter) with cutoff frequency $f_0$ = 50 MHz and time window of the gated integration $t$ = 160 ns were applied, in which the characteristic value was extracted from the averaged waveform of 2000 events. The changes in the noise factors with the sampling frequency are shown in Fig. 5.

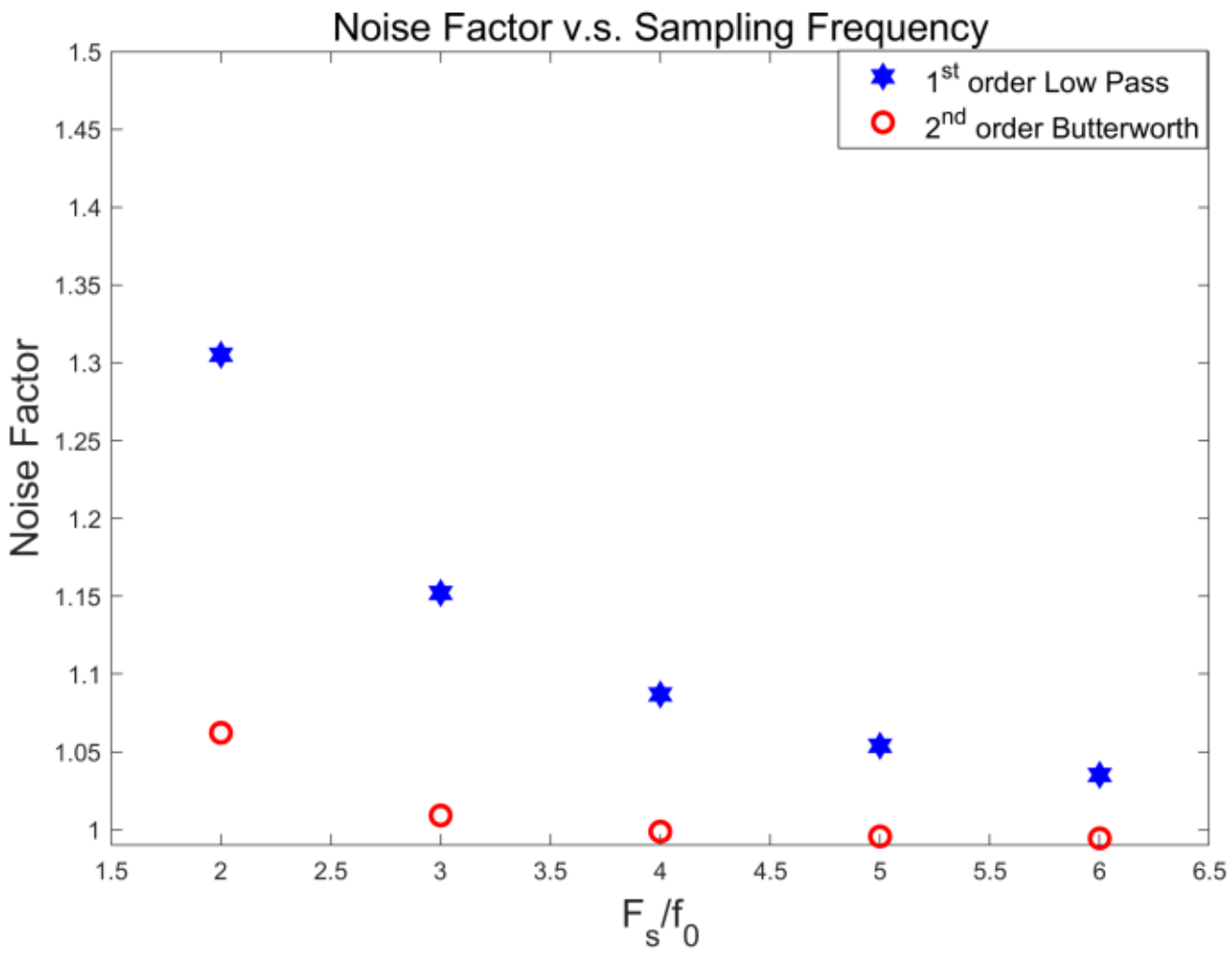

Fig. 5 Noise factor influenced by sampling frequency

From Fig. 5, we can conclude that the band-limited noise of the fast amplifier is determined by the power spectral density of the input noise as Eq. (20) as long as the sampling rate is sufficiently high (> 3-4 $f_0$). This result is in accordance with the results obtained by Can Liao (2015) [6] in which the sampling rate does not significantly affect the PSD quality when it is sufficiently high to capture enough pulse shape information. Generally, a second-order digital filter can be applied for pre-processing of the digital signal to minimise the degradation due to the limit of the sampling rate.

*Summary of Gated-Integration Noise Analysis in the Digital Domain*

According to Eqs. (16) and (20), if the assumptions are satisfied, namely,

1. The quantisation error of ADC is uncorrelated, which is satisfied when the input-referred noise is larger than one-half LSB or the ac signal spans more than a few LSBs;
2. Gated integration time $t \gg 1/\omega_0$, where $\omega_0$ can be chosen according to the highest frequency of the signal $f_0$, i.e. $\omega_0 = 2\pi f_0$;
3. The sampling rate is sufficiently high to reduce the noise factor (normally 3–4 $f_0$)

then, the uncertainty of the digital gated charge integration caused by the noise ends up as

$$\begin{aligned}\sigma_{Q_n}^2 &= \frac{1}{2} S_{x0} \cdot t = \frac{1}{2}\left(S_{x_I} + S_{x_ADC}\right) \cdot t \\ &= \frac{1}{2}\left(S_{x_I} + \frac{V_{n_ADC}^2}{A^2 \cdot F_s / 2}\right) t\end{aligned} \tag{22}$$

where $S_{x0}$ is the power spectral density of the total input-referred noise, including those of the fast amplifier and ADC. The influence of Eq. (22) on both energy resolution and PSD performance will be discussed in detail in Sections 4.1 and 5.1.

# 3. Experimental Setup

In this research, a 2 × 2 in cylindrical $LaBr_3$:Ce detector was used, which is commercially available from Saint-Gobain with a spectral optimised R6233-100 PMT. A 12-bit, 2.5-Gs/s Lecroy oscilloscope HDO6104 was used as a digitiser. Sources $^{137}Cs$ and $^{22}Na$ were used for energy calibration, and the resolution of the full peak of $^{137}Cs$ at 662 keV was chosen for experimental verification. Meanwhile, environmental background measurement was made for the PSD estimation.

Two setups, namely, with and without a fast amplifier, were built to separately verify the influence of a fast amplifier (correlated noise) and the ADC (uncorrelated noise).

a) Setup 1 (without a fast amplifier): the PMT output was directly connected to a 50-Ω input of the oscilloscope via a coaxial cable, in which the pulse amplitude of 662keV gamma ray is 44mV (41.6pC charge). The full scale was set to $400 \text{ mV} = 8 \text{ div} \times 50 \text{ mV/div}$.
b) Setup 2 (with a fast amplifier): a current-to-voltage converter amplifier with a gain of 250 Ω and bandwidth of 350 MHz was used as the fast amplifier (shown in Fig. 8), in which the amplitude of 662keV gamma ray is 220mV. The full scale was set to $800 \text{ mV} = 8 \text{ div} \times 100 \text{ mV/div}$.

The application of fast Fourier transform analysis to the averaged $LaBr_3$ waveforms shows that its highest frequency is approximately 50 MHz, at which point its amplitude in the frequency domain decreases 40dB. Therefore, the sampling rate must be greater than 100 Ms/s according to the Nyquist theorem.

The pre-processing actions, which were aimed at better reconstruction and alignment of the input signals, are listed below:

a) The waveforms were reconstructed using spline interpolation (applied with ROOT data analysis framework) for low sampling-rate digitisation to obtain a better precision while using discrete sum to substitute the analogue integration.

b) A 100-MHz second-order, digital low-pass filter was applied to reduce the noise, and the pulses were aligned using the 20% constant fraction timing method to obtain a better alignment.

# 4. Optimal Design of Waveform Digitisers for Energy Resolution

*4.1 Quantitative Analysis of Energy Resolution*

In general, the measured energy resolution consists of the intrinsic resolution caused by statistical fluctuation and the uncertainty caused by noise due to electronics.

$$\begin{aligned} res_{\mathrm{m}}^2 &= res_{\mathrm{intrinsic}}^2 + res_{\mathrm{noise}}^2 \\ &= res_{\mathrm{intrinsic}}^2 + \left(\frac{2.355\sigma_{Q_n}}{Q_t}\right)^2 \end{aligned} \tag{23}$$

We have known that a charge-sensitive amplifier is usually applied on semiconductor detectors owing to its low noise, which can be considered negligible for spectral system based on scintillators. In our experiment, the intrinsic energy resolution of $LaBr_3(Ce)$ is approximated by measurement using a charge sensitive preamplifier and a multi-channel analyser, namely, 2.62% at 662 keV. Meanwhile, $Q_t$ can be extracted from the mean value of the Gaussian peak from the integrated charge spectrum, that is, 41.6 pC at 662 keV.

Fig. 2 shows that the long gate integration corresponds to the total deposited charge. Attention has to be paid that the baseline should be properly subtracted. The integrated noise can be shown as follows:

$$Q_n = -T_s \cdot \left( \sum_{j=t_0}^{t_0+t_T} i_j - \frac{t_T}{t_B} \sum_{k=t_0-t_B}^{t_0} i_k \right), \tag{24}$$

where $t_0$ is the starting time of the input pulse, $t_B$ is the time window of the baseline gate, and $t_T$ is the time window of the long gate. Thus, the uncertainty of the integrated noise can be expressed as

$$\sigma_{Q_n}^2 = \frac{1}{2} S_{x0} \cdot \left(1 + \frac{t_T}{t_s}\right) \cdot t_T . \tag{25}$$

By substituting Eq. (25) to Eq. (23), the measured energy resolution can be calculated as

$$res^2 = res_{\mathrm{intrinsic}}^2 + \left(\frac{2.355}{\mathrm{Q}_t}\right)^2 \cdot \frac{1}{2} S_{x0} \cdot \left(1 + \frac{t_T}{t_B}\right) \cdot t_T . \tag{26}$$

According to Eq. (22), the noise power spectral density is influenced by the fast amplifier, sampling rate, and vertical resolution. In Sections 4.2 and 4.3, we will present the analysis of the influences of these factors in detail using several experiments.

*4.2 Verification and Optimisation of Setup 1 (Without a Fast Amplifier) for Energy Resolution*

By considering the full-energy peak of $^{137}$Cs at 662 keV as a representative, the measurable parameters contained in Eq. (26) are as follows: *I–V* gain *A* = 50 Ω, the root mean square (rms) noise of the ADC measured from baseline is $V_{n_ADC} = 2.90$ mV, and gated information $t_T = t_B = 160$ ns. On the basis of these parameters, theoretical prediction of the energy resolution under digital systems with different characteristics can be calculated using Eq. (26).

*Influence of Sampling Rate on Energy Resolution*

For Setup 1 without a fast amplifier, the original data were obtained at a 2.5-Gs/s sampling rate. Offline down-sampling data processing was performed under different sampling rates from 100 Ms/s to 2.5 Gs/s to study the influence of sampling rate on the energy resolution. The experimental results well coincided with the theoretical prediction, as shown in Fig. 6.

Fig. 6 Energy resolution of the $^{137}$Cs peak at 662 keV that varies with sampling frequency

*Influence of Vertical Resolution on the Energy Resolution*

For Setup 1, *FUS* = 400 mV, *N* = 12 bit, and the noise caused by quantisation error is 0.028 mV [Eq. (11)], which is negligible compared with the measured 0.29 mV (ENOB = 8.64) mainly caused by the input-referred noise.

$$V_{ir} = \sqrt{V_{n_ADC}^2 - V_q^2} \approx 0.289mV \tag{27}$$

Research about lowering the vertical resolution of the ADC is performed through offline processing. To be specific, the noise from different origins related to the vertical resolution is listed in Table 1, and the energy resolution that changes with the vertical resolution is shown in Fig. 7, where the theoretical prediction is indicated by a red line.

Table 1. Noise contribution of the input-referred noise and quantisation error

| Bit resolution | 12 | 11 | 10 | 9 | 8 | 7 |
|---|---|---|---|---|---|---|
| $V_q$/mV | 0.028 | 0.056 | 0.113 | 0.226 | 0.451 | 0.902 |
| $V_{ir}$/mV | ***0.289*** | | | | | |
| $V_{n_ADC}$/mV | 0.290 | 0.294 | 0.310 | 0.366 | 0.535 | 0.947 |
| $LSB$/mV | 0.098 | 0.195 | 0.391 | ***0.781*** | 1.563 | 3.125 |

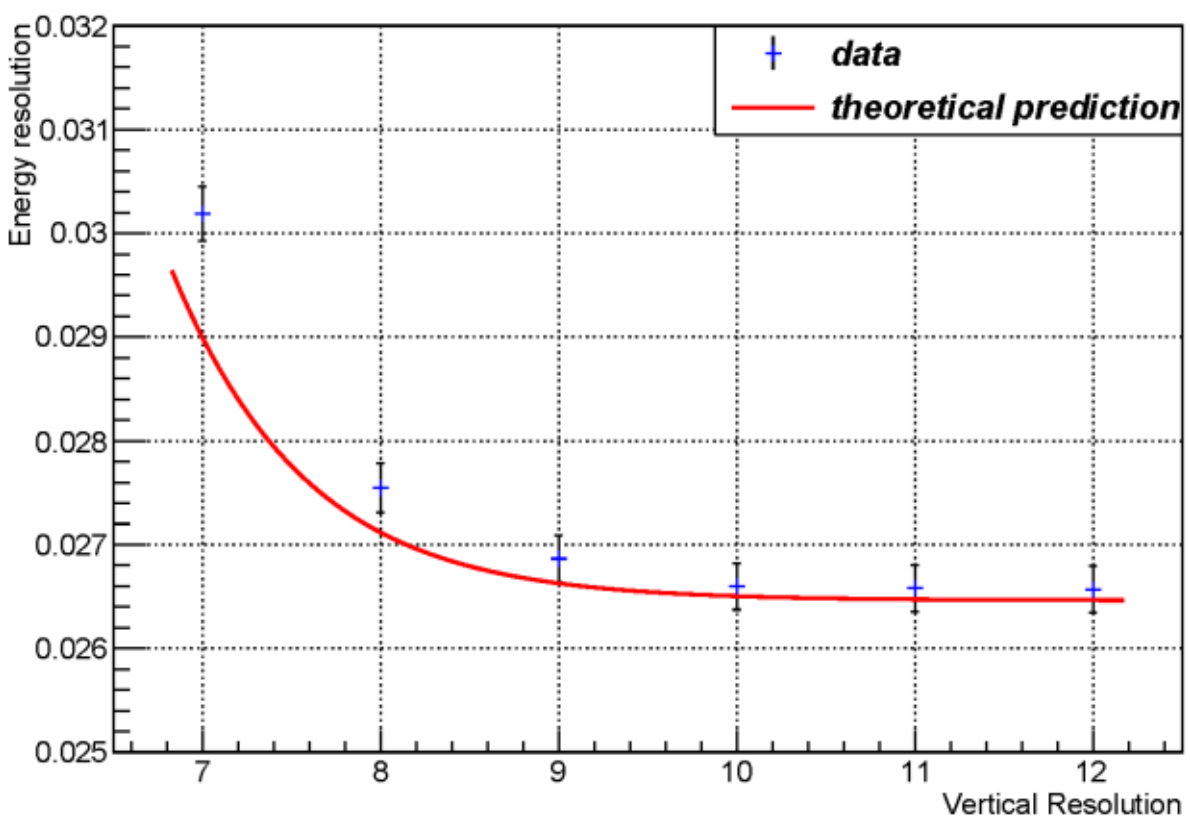

Fig. 7 Energy resolution versus vertical resolution of the ADC

According to the assumptions presented in the summary in Section 2, if the bit resolution is less than 9 bits, the *input-referred* noise *(*0.289 mV*)* is less than one-half LSB (0.781mV/2). In particular, when the bit resolution reduces to 7 bit, the ac signal at 662 keV only spans to approximately 14 LSBs. The quantisation error will no longer be uncorrelated, and the uncertainty of the gated integration of the noise will be larger. Therefore, the experimental energy resolution is worse than the theoretically predicted value indicated by the red line.

To conclude, we need to be very careful when the sampling resolution is too low and the ac signal is too small to span a few LSBs, which is rare and should be avoided in real applications.

*4.3 Verification and Optimisation of Setup 2 (with fast amplifier) for Energy Resolution*

In Setup 2, a fast amplifier (see Fig. 8) with *I–V* gain $A = 250\ \Omega$ and bandwidth of 350 MHz was applied. The input-referred noise power spectral density of the fast amplifier is expressed as

$$S_{x_I} \approx I_{BN}^2 + \frac{E_{NI}^2}{R_S^2} + \frac{4kT}{R_S} + \frac{I_{BI}^2}{R_S^2} R_G^2 + \frac{4kT}{R_S^2} R_G + \frac{4kTR_G^2}{R_F R_s^2}\ . \tag{28}$$

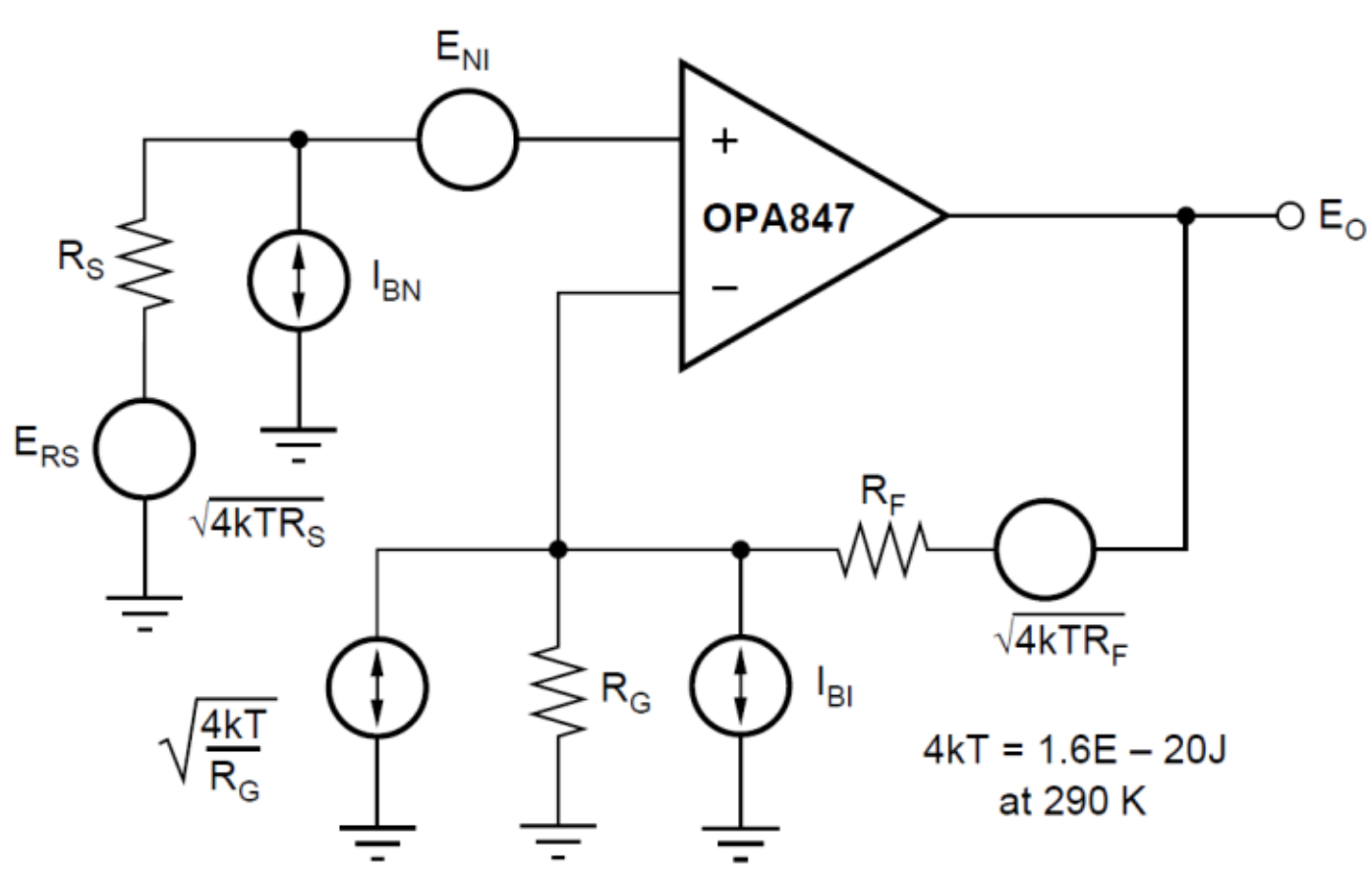

Fig. 8 Noise sources of the fast amplifier

According to the datasheet of integrated chip OPA847 with parameters $R_S = 25\ \Omega, R_G = 39.2\ \Omega, R_F = 750\ \Omega,$ and $R_O = 50\ \Omega,$ the input-referred noise power spectral density at 300 K is $2.81 \times 10^{-21} A^2 / Hz$. According to Eq. (26), the contribution of fast amplifier to the resolution is:

$$res_{FA} = \sqrt{\left(\frac{2.355}{Q_t}\right)^2 \cdot \frac{1}{2} S_{x_I} \cdot \left(1 + \frac{t_T}{t_B}\right) \cdot t_T} = 0.12\% \tag{29}$$

The influence of fast amplifier on energy resolution is negligible owing to the fact that $0.12\% \ll 2.62\%$.

For the ADC part, the measured rms noise of the ADC is $V_{n_ADC} = 0.5$ mV for 100 mV/DIV, and the *I–V* gain of the system is 250 Ω. The energy resolution changes with the sampling rate, and the theoretical prediction is shown in Fig. 9.

Fig. 9 Energy resolution of the $^{137}$Cs peak at 662 keV with a fast amplifier, which varies with the sampling rate

Both the calculation and offline experiment show that the amplification of the fast amplifier can improve the SNR of the ADC by making full use of the ADC range, especially for a normal ADC with a fixed full scale. Furthermore, as a design guide of a fast amplifier, the rule of thumb is to reduce the power spectral density of the input noise, which can be optimised using Eq. (28) by increasing $R_S$ and reducing $R_G$ in this setup.

*4.4 Summary of Energy Resolution for LaBr(Ce) Detector*

For Setup1, to ensure that the energy resolution is better than 2.8% at 662 keV, while the pulse amplitude is 44 mV at 50 Ω. As shown in Fig. 6, the characteristics of the digitiser should be at least $Fs$ > 350 Ms/s and $V_{n_ADC} < 0.29\ mV$ (*ENOB*> 8.6, *FUS*=400mV). The increase in the sampling rate as a factor of four can be achieved at the expense of decreasing 1 ENOB based on Eq. (22), where no fast amplifier is contained($S_{x_I} = 0$).

For Setup 2, The fast amplifier made larger pulse amplitude and full use of the ADC range (800mV), by amplifying the amplitude of gamma ray at 662keV from 44mV (at 50 Ω) to 220mV, which results in a better energy resolution as shown in Fig. 9.

# 5. Optimal Design of Waveform Digitisers for PSD

*5.1 Quantitative Analysis of the PSD Feature*

According to Eq. (1), the PSD feature of the CCM is defined as the ratio between the partial and total charges. The main part of the feature extraction is also the gated integration. The above analysis can also be applied to PSD analysis.

The distribution changes with the ADC properties. As an example, the CCM feature distributions at a sampling rate of 0.1 and 2.5 Gs/s are shown in Fig. 10.

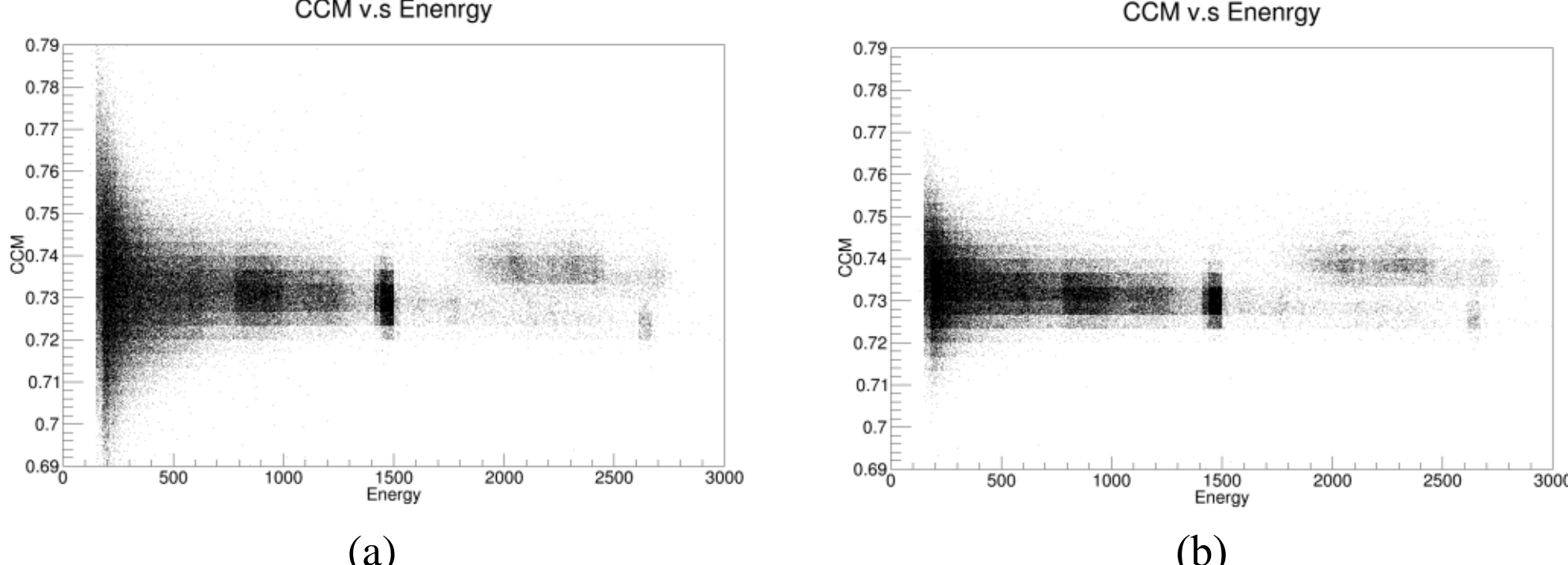

(a) (b)

Fig. 10 CCM distribution versus energy at (a) 0.1 Gs/s and (b) 2.5 Gs/s

Fig. 11 shows that the mean value of the CCM remains the same, which is determined by the detector properties, but the distribution spreads more at lower sampling rates. According to our previous work [2], the uncertainty of the CCM mainly consists of the intrinsic uncertainty and noise, which are respectively proportional to $1/\sqrt{Q}$ and $1/\mathrm{Q}$.

$$\begin{aligned}\sigma_{CCM}^{2} &= \sigma_{CCM_Intrinsic}^{2} + \sigma_{CCM_Noise}^{2} + \sigma_{CCM_\mathrm{Jitter}}^{2} \\ &\approx \left(\frac{c1}{\sqrt{Q}}\right)^{2} + \left(\frac{c2}{Q}\right)^{2}\end{aligned} \quad . \qquad (30)$$

According to the definition of Eq. (1), the CCM uncertainty can be calculated as

$$\sigma_{CCM}^{2} = \left(\frac{\partial CCM}{\partial Q_{p}}\right)^{2} \sigma_{Q_p}^{2} + \left(\frac{\partial CCM}{\partial Q_{t}}\right)^{2} \sigma_{Q_t}^{2} + 2\frac{\partial CCM}{\partial Q_{p}}\frac{\partial CCM}{\partial Q_{t}} \operatorname{cov}\left[Q_{p}, Q_{t}\right] \qquad (31)$$

Here, we only analyse the uncertainty caused by the noise.

$$Q_{p_n} = -T_{s} \cdot \left( \sum_{j=t_1}^{t_1+t_P} i_{n}(j) - \frac{t_P}{t_B} \sum_{k=t_0-t_B}^{t_0} i_{n}(k) \right) \qquad (32)$$

$$\sigma_{Q_{P_n}}^{2} = \frac{1}{2} S_{x0} \cdot \left(1 + \frac{t_P}{t_B}\right) \cdot t_P \qquad (33)$$

The partial charge integration window is a part of the total charge integration window. The correlation between $Q_{p_n}$ and $Q_{t_n}$ is then obtained as follows:

$$\begin{aligned}\operatorname{cov}\left[Q_{p_n}, Q_{t_n}\right] &= T_{s}^{2} \cdot \operatorname{cov}\left[ \sum_{j=t_1}^{t_1+t_P} i_{n}(j) - \frac{t_P}{t_B} \sum_{k=t_0-t_B}^{t_0} i_{n}(k), \sum_{j=t_0}^{t_0+t_T} i_{n}(j) - \frac{t_T}{t_b} \sum_{k=t_0-t_B}^{t_0} i_{n}(k) \right] \\ &= \frac{1}{2} S_{x0} \cdot \left( t_P + \frac{t_P t_T}{t_B} \right)\end{aligned} \quad . \qquad (34)$$

Therefore, the uncertainty of the CCM caused by the noise can be concluded as

$$\sigma^2_{CCM_Noise} = \frac{1}{2}\frac{S_{x0}}{Q_t^2}\left[\left(1+\frac{t_P}{t_B}\right)t_P + \overline{CCM}^2\left(1+\frac{t_T}{t_B}\right)t_T - 2\overline{CCM}\left(t_P+\frac{t_P t_T}{t_B}\right)\right]. \quad (35)$$

Substituting parameters $t_p = 42.8, t_T = t_B = 160$ ns, and $\overline{CCM} = 0.73$ to the above equation yields

$$t_{eff} = \left(1+\frac{t_P}{t_B}\right)t_P + CCM^2\left(1+\frac{t_T}{t_B}\right)t_T - 2CCM\left(t_P+\frac{t_P t_T}{t_B}\right) \approx 100ns. \quad (36)$$

Eq. (30) can be replaced by

$$\sigma^2_{CCM} \approx \left(\frac{c1}{\sqrt{Q_t}}\right)^2 + \frac{1}{2}\frac{S_{x0}\cdot t_{eff}}{Q_t^2}. \quad (37)$$

Similar to Eq. (22), the CCM is also influenced by the noise power spectral density ($S_{x0}$). The influence of the sampling rate and vertical resolution on PSD is similar to the detailed analysis in the optimal design for energy resolution. Therefore, for verification, in Section 5.2, we present the fitting parameters of $c1$ and $S_{x0}$ (compared with the measured value) using the data from Setup 1. Hence, the prediction of the CCM uncertainty versus sampling rate using Eq. (37) is also verified in Section 5.3.

*5.2 Fitting of Intrinsic and Noise Contribution to the CCM Uncertainty*

According to Eq. (37), the uncertainty of the CCM feature is composed of the intrinsic statistical fluctuation and noise uncertainty. Let $c1$ and $V_{n_ADC}$ (i.e. $S_{x0}$) be two free parameters. The fitting results for Setup 1 are shown in Fig. 11. Fitting was applied using the sampling rate at 0.25 Gs/s in which the uncertainties of the CCM caused by these two parts are comparable. The fitting parameters are expressed as

$$\begin{aligned} c1 \quad &= 0.0226 \pm 0.0001 \\ V_{n_ADC} &= (0.295 \pm 0.001)\,mV \end{aligned}. \quad (38)$$

The fitted rms noise of the ADC is $0.295 \pm 0.001$mV, which coincides well with the measured value of 0.290 mV.

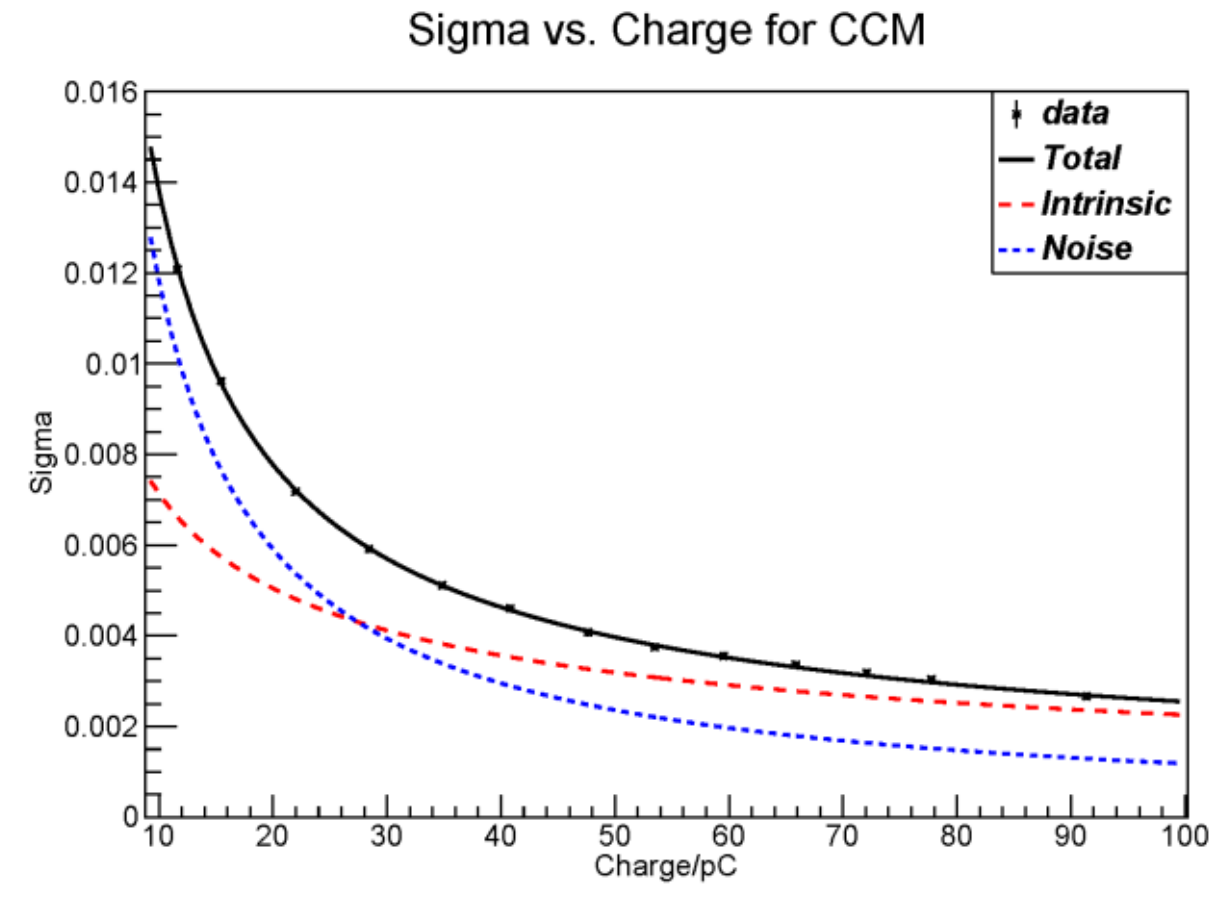

Fig. 11 Fitting of the CCM uncertainty with a sampling rate of 0.25 Gs/s

*5.3 Prediction of the CCM Uncertainty with Different Sampling Rates*

According to the fitting results presented in Section 5.2, we can predict the uncertainty of the CCM with different digitiser properties using Eq. (37). The distribution of PSD feature, for a fixed energy 662keV (Q=41.6pC), is shown as Fig. 12.

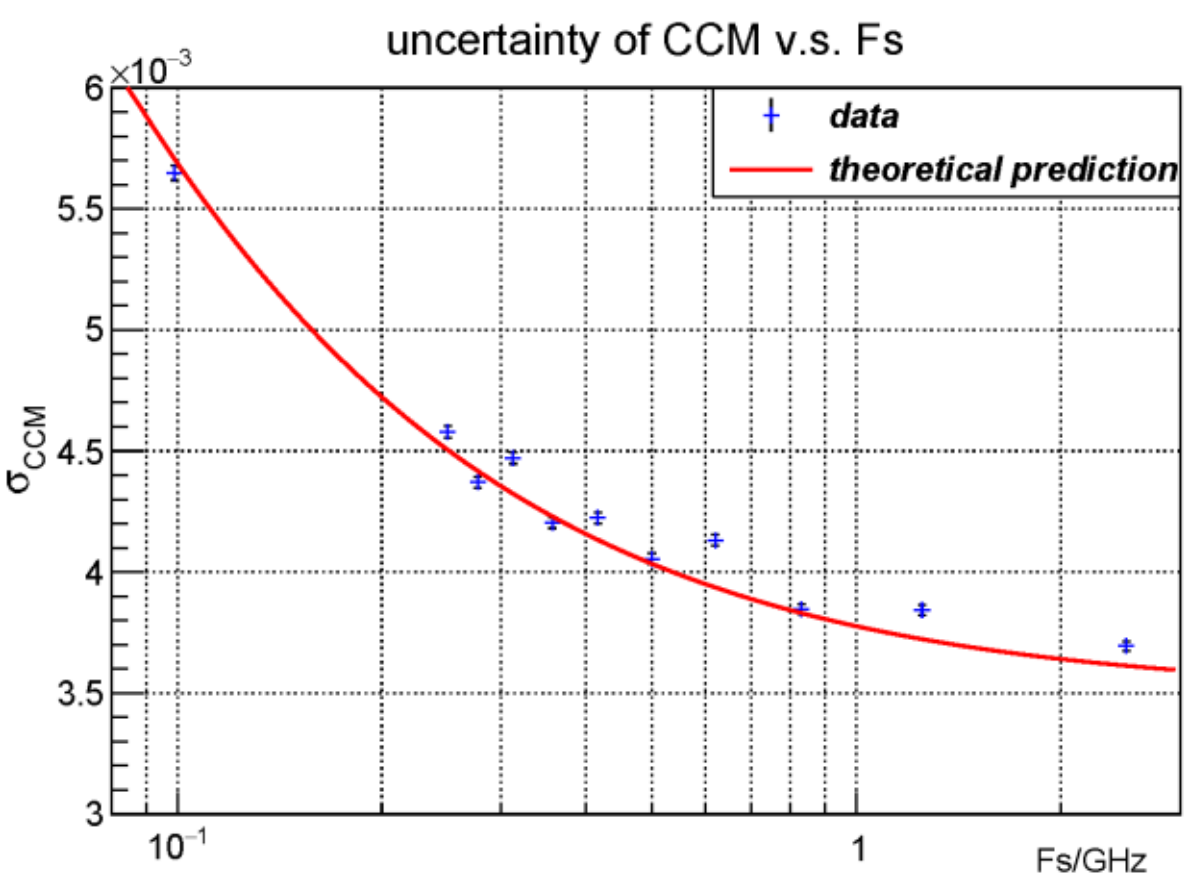

Fig. 12 Uncertainty of the CCM versus sampling rate

Fig. 12 shows that the theoretical prediction of the uncertainty of the CCM versus sampling rate fits the data well. That is, once the intrinsic uncertainty of the CCM is achieved from one fitting measurement using Eq. (37), the spreading of the PSD feature with different digitiser properties can also be calculated using Eq. (37).

## 6. Conclusion

In this study, the noise model of a time-variant gated integration for a waveform digitiser has been quantitatively calculated using Eq. (22), which is suitable for the analysis of both energy resolution and PSD application. The energy resolution of a waveform digitiser system can be estimated using Eqs. (22) and (26), and the spreading of PSD feature can be estimated using Eqs. (22) and (37).

On the basis of the model, as illustrated in section 4 and 5, the influences of the waveform digitiser properties in terms of fast amplifier, sampling rate, and vertical resolution on both energy resolution and PSD performance are discussed and verified separately by experiments, using $LaBr_3(Ce)$ detector.

To conclude, the estimation of the optimal design for waveform digitisers based on the above analysis can also be generally applied to other scintillators or similar conditions based on the pulse shape analysis.

## Acknowledgements

This work was supported by the Tsinghua University Initiative Scientific Research Program (Grant No. 2014Z21016).